# A Probit Estimation of Urban Bases of Environmental Awareness: Evidence from Sylhet City, Bangladesh.

Mohammad Masud Alam[1] ; AFM Zakaria[2]

## Abstract

This paper evaluates the significant factors contributing to environmental awareness among individuals living in the urban area of Sylhet, Bangladesh.  Ordered Probit(OPM) estimation is applied on the value of ten measures of individual environmental concern. The estimated results of OPM reveal the dominance of higher education, higher income, and full-employment status on environmental concern and environmentally responsible behavior. Younger and more educated respondents tended to be more knowledgeable and concerned than older and less educated respondents. The marginal effect of household size, middle-income level income, and part-time employment status of the survey respondents played a less significant role in the degree of environmental awareness. Findings also validate the "age hypothesis" proposed by Van Liere and Dunlap (1980), and the gender effect reveals an insignificant role in determining the degree of environmental concern. Environmental awareness among urban individuals with higher income increased linearly with environmental awareness programs which may have significant policy importance, such as environmental awareness program for old-aged and less-educated individuals, and may lead to increased taxation on higher income group to mitigate city areas' pollution problems.

[1] Department of Economics, Shahjalal University of Science & Technology, Sylhet. Email:masud.sust@gmail.com
[2] Department of Anthropology, Shahjalal University of Science & Technology, Sylhet.

## Introduction

In the era of globalization, environmental issues become the core in the scholarships of social sciences due to their inseparable relationships between economy, environment, and how citizens interact with the environment. An economic plan is frequently motivated by factors such as the green environment and sectoral economic policy variables (Markandya et al., 2002). Environmental concern among cities and city populations can shape the country's environmental future, including environmental governance lace with individual perception and socioeconomic determinants of environmentally responsible behavior. The economic dimension of urban development plans (Roberts, and Grimes, 1997) is also incomplete if any inattention of environmental protection and without resorting to legal frameworks and economic instruments of environmental or resource factors ((Sigeki, Reeitsu, Hideo, and Shigeki, 1997).

In recent years research interest has grown in the protection of the environment in the city areas of Bangladesh due to its major social, political, and economic importance for ensuring the long-term sustainability in improving environmental quality. Researchers and academics interested in protecting the environment have commonly documented that personal and social awareness and subsequent concern or true preferences are at the heart of environmental protection (Duroy, 2005; Shen and Saijob, 2007; Liere and Dunlap, 1980; Wall, 1995). This study is an attempt to investigate the influence of social and economic characteristics on different measures of environmental concern of individuals in Sylhet, Bangladesh. It attempts to clarify how environmental awareness can be utilized as a tool for environmental policymaking and management. Issues that policymakers want to know the determinants of social and urban bases of environmental concern so that regulatory attempts and social initiatives can provide the optimum outcome for environmental protection.

This paper sets the core objective of finding out the correlations between these determinants and the environmental concern of urban populations. Respondent's true preference is one of the most pertinent information to take essential decisions on taxation, provision of public good for urban people, and pricing of public utilities. Therefore research findings will be an important step for policymakers to clarify the relationship between environmental quality and well-being and perhaps give influential ways to control environmental degradation.

In terms of the organization of this paper, the next section outlines the theoretical motivation with a varied array of environmental issues, urbanization, and hypotheses examined in the previous studies. Section three presents a brief outline of data and variables. We introduce the model specification in section four and analysis of empirical results in section five. Finally, section six presents the policy implications of findings, their limitations, and concluding remarks with future research direction.

## Theoretical Motivation

The term environmental concern in this study mainly originated from the definition presented by Jones and Dunlap (2002): "…the degree to which people are aware of problems regarding the environment and support efforts to solve them and indicate a willingness to contribute personally to their solution". Environmental concern indicates the degree to which people are aware of the environment and support efforts to solve them and demonstrates the willingness to contribute personally to their solution. Stern and Dietz (1994) define environmental concern as egoistic, social-altruistic, and biospheric value orientations and beliefs about the consequences of environmental changes for valued objects. Urbanization refers to an expansion of economic amenities in the city areas with modern and technological facilities. Urbanization is a process of the growth of cities, increased industrialization, development, which leads to changes in specialization, labor division, and human behaviors.

Most of the research and theory on environmental concern includes attitudinal studies, experimental and quasi-experimental surveys, and applied research on environmental attitudes and behaviors (Buttel, 1987). Research on attitude theory (Bagozzi and Warshaw 1990, Harry 1971, Maslow 1970), human behavior (Howell and Laska, 1992, Stern, Dietz, and Kalof, 1993), and environment and behavior (Poortinga, Steg, and Vlek, 2004, Schahn and Hotzer, 1990) envisage environmental concern as an evaluation of an individual to the environment or attitude towards facts or own individual behavior to the environmental protection.

Two influential studies have shown how age, gender, social class, household size, and political ideology can explain various environmental attitudes and behaviors (Van Liere and Dunlap 1980, Fransson and Garling, 1999). The age hypothesis documents the inverse relationship between age and environmental concern and settles the proposition that younger people tend to be more concerned about environmental protection than elders. Considering gender impact on environmental concern, women appear to be slightly and consistently more concerned about the environment than males (e.g., Davidson and Freudenburg 1996). Education, occupational status, and income are positively related to environmental concerns, as demonstrated by the social class hypothesis (Andrews and Stephan, 1978; Inglehart, 1990; Schaha and Holzer 1990; Van Liere and Dunlap, 1980). Regarding the residence hypothesis, those living in metropolitan areas were significantly more concerned than those living in towns or the countryside (Fransson and Gorling 1999). Referring to U.S. data, the link between political ideology and environmental concerns shows that people who support Democrats and liberals are more concerned about environmental quality than those who support republican and conservative counterparts (Buttel and Flinn 1978, Van Liere and Dunlap 1980).

**Data**

Three hundred twenty individuals from different locations in Sylhet city, Bangladesh, participated in the survey and completed the questionnaire in the fall of 2012. The questionnaire consists of 34 questions related to significant environmental issues. The target population is people with different backgrounds, including gender, age, education levels, location of residency, and social status. The information was collected according to the population density throughout five major areas of city areas. **Table I** shows the summary information of all the associated socioeconomic characteristics of the data. Following Shen and Saijo (2007), we have set one index to estimate the specification for general environmental concern, used eight indices for specific environmental issues including land, energy, and environmental issues of Sylhet city area, and finally one index to examine the attitude towards pro-environmental behavior.

Independent Variables

The independent variables included in this study were six demographic, socioeconomic, and regional factors. A set of dummy variables were developed.

1. Gender. Female was coded (1), and male, the reference group, was coded (0).

2. Social class. Six dummy variables were developed for social class: (a) Higher education was coded (1), if respondent completed undergraduate level and bellow undergraduate, the reference group, was coded (0). (b) Higher income class was coded (1) when monthly income level is more than BDT 100,000, otherwise (0) and the middle-income class was coded (1) if monthly income in between 60,000-99,999 BDT. (c) Full employment was coded (1) if the respondent is fully employed, otherwise (0), and part-time employed respondent coded (1), otherwise (0).

**Table I:** summary statistics of independent variables.

| Variable | Mean | S.D. |
|---|---|---|
| Female (=1 if female) | 0.62 | 0.48 |
| Age: actual age of respondent | 32.15 | 9.72 |
| High_ed (=1 if above undergraduate) | 0.47 | 0.49 |
| High_inc(=1 if income above 100,000) | 0.52 | 0.49 |
| Low_inc(=1 if income in between 20000-50000) | 0.27 | 0.44 |
| House_size: Actual household size | 5.57 | 1.74 |
| Ful_emp(=1 if respondent is full employed) | 0.48 | 0.50 |
| Prt_emp(=1 if respondent is part time employed) | 0.38 | 0.48 |

We provide a correlation matrix (in appendix) that reveals whether any significant correlation among all the independent variables. It displays moderate negative correlations between age and part-time employment, high income and middle income, full employment, and part-time employment. Separate regression models excluding two collinear variables have been estimated to avoid this multicollinearity problem. All other variables exhibit low correlation, and therefore co-linearity is not a problem in the regression analysis.

Dependent Variable – The dependent variables in this study are a general environmental concern: concern about global warming, concern about city areas and cross-areas pollution, concern about air/water/soil problem, concern about energy problem, Sylhet city green land, an ecological problem, concern about a health problem, concern about recycling, concern about noise/odor and environmental consideration when buying electronics.   All variables are measured by Likert type response formats, using a five-point response anchored: 'not concern,'

'not quite concern' 'neither concern nor unconcern,' 'somewhat concern,' and 'concern.' These five responses correspond to censoring values 0, 1, 2, 3, and 4, respectively. The higher score indicated more favorable environmental attitudes & stronger norms. Variable VJ examines pro-environmental behavior where respondents were asked to rank environmental considerations when buying electronics. The scores of this statement were coded as: consider environmental impact as the first issue (4) second issue (3) third issue (2), and neither (1). Table II shows these variables that we have used in the questionnaire to demonstrate the different measures of individual environmental concern.

**Table II:** Summary statistics of environmental concern indices

| Description | Mean | (S.D.) |
|---|---|---|
| VA: Concern about general environmental problem | 3.44 | 0.81 |
| VB: Concern about global warming problem | 3.15 | 1.15 |
| VC: Concern about City areas pollution and cross-areas pollution problem | 3.14 | 1.50 |
| VD: Concern about air/water/soil pollution problem | 3.36 | 0.87 |
| VE: Concern about the energy problem | 2.97 | 1.10 |
| VF: Concern about Sylhet City's green land and ecological problems | 2.95 | 2.01 |
| VG: Concern about the effect of harmful substances on health | 2.76 | 1.09 |
| VH: Concern about disposal, reduction, and recycling of waste | 2.45 | 1.17 |
| VI: Concern about living environmental problems such as noise/odor | 3.23 | 0.86 |
| VJ: Rank of environmental consideration when buying electronics | 2.24 | 1.62 |

## Model specification

The study employed the ordered probit model (OPM) in which Likert-type response formats measured environmental concerns. Because of the discrete nature of the dependent variable in this study, ordinary least squares regression would be an inappropriate model. Through the ordered probit model, we identify which socioeconomic factor determines individual environmental concern and discern which factor plays the most important role by predicting each factor's marginal effect on the probability of being the most and least environmentally concerned individual (**Shena and Saijo, 2007**). We can describe the choices of concern as a discrete variable, Yi. A respondent's probability of falling into one of the five environmental concerns categories can be decomposed into a deterministic component and an additive stochastic component. According to our categorization, this variable can take one of the following five values

$y_i = 0$ , if the degree of concern is chosen as 'Not concern'

$y_i = 1$, if the degree of concern is chosen as 'Not quite concern'

$y_i = 2$, if the degree of concern is chosen as 'Neither concern nor unconcern'

$y_i = 3$, if the degree of concern is chosen as 'Somewhat concern'

$y_i = 4$, if the degree of concern is chosen as 'Concern'

Then, we assume that individual environmental concern is based on a continuous and latent variable $y*_i$. This latent variable is assumed to be a linear function of all the socioeconomic variables, and the standard ordered probit model is widely used to analyze discrete data of this variety and is built around a latent regression of the following form:

$$y_i^* = \beta'X_i + \varepsilon_i \text{ for i} = 1,2,\ldots\ldots\ldots,N \qquad (i)$$

Where $y*_i$ is the unobserved measure of environmental concern, $X_i$ is a vector of independent variables, N is the number of respondents, and $\varepsilon_i$ is the error term. The relationship between $y*_i$ and the ordinal likelihood ranking (observed category) y is a function of the cut-off points or thresholds, $k_1$s, which are unknown but can be estimated along with $\beta s$.

Let $k_1 \prec k_2 \prec k_3 \prec k_4$ be unknown cut points or threshold parameters and with zero as a normalization assumption, the censored variable can be define as:

$y_i = 0$, if $y_i^* \leq k_1$ (ii)

$y_i = 1$, if $k_1 \prec y_i^* \leq k_2$ (iii)

$y_i = 2$, if $k_2 \prec y_i^* \leq k_3$ (iv)

$y_i = 3$, if $k_3 \prec y_i^* \leq k_4$ (v)

$y_i = 4$, if $y_i^* \succ k_4$ (vi)

Note that the four cut points are estimated along with the coefficients of the independent variables of vector Xi .Consequently, the probabilities of $y_i$ being classified as 'not concern', 'not quite concern' 'neither concern nor unconcern', 'somewhat concern', and 'concern' are given by

$\Pr(y_i = 0) = \Pr(\beta'X_i + \varepsilon_i \leq k_1)$ (vii)

$\Pr(y_i = 1) = \Pr(k_1 \prec \beta'X_i + \varepsilon_i \leq k_2)$ (viii)

$\Pr(y_i = 2) = \Pr(k_2 \prec \beta'X_i + \varepsilon_i \leq k_3)$ (ix)

$\Pr(y_i = 3) = \Pr(k_3 \prec \beta'X_i + \varepsilon_i \leq k_4)$ (x)

$\Pr(y_i = 4) = \Pr(\beta'X_i + \varepsilon_i \succ k_4)$ (xi)

With a normal distribution of the error term ε in equation (vii) - (xi), the cut points κ and coefficients β can be estimated as an ordered probit model (OPM). The estimation of OPM is based on the maximization of the likelihood function, which is expressed as:

$$\log L = \sum_{1}^{j} \log \Pr(y = j/x)$$

In terms of available data for this study, latent regression can be formulated as:

$$y_i = \beta_0 + \beta_1 male + \beta_2 age + \beta_3 high\_ed + \beta_4 income(high, low) +$$
$$\beta_5 occupation(Full\_emp, \Pr t\_emp) + \beta_6 house\_size + \varepsilon_{io}............................(xii)$$

Results for each model will be presented in the next section, along with an explanation. From the slope parameter and threshold estimates, it is relatively straightforward to calculate individuals' five choices of environmental concern with their marginal effects. Given the cumulative normal function, φ (β′x), and the probabilities, we can estimate the associated marginal impact and examine in what direction those effects are exerted.

**Analysis of Results**

Empirical results of this OPM technique are presented in terms of a specific environmental choice of concern, in which choice of concerns are classified into five categories. Two separate regression models have been estimated: one model includes basic eight variables, and another model excludes two collinear variables such as mid-level income and part-time employment status. The OPM estimates of both models' eight and six slope coefficients are reported in Table III – XII in the appendix. The unknown threshold parameters are not reported because of space limitations. Cut points or threshold parameters are estimated by the data and help to match the probabilities associated with each discrete outcome. Note that the estimated coefficient has no direct interpretation but can calculate probabilities of getting a different choice of concerns and their corresponding marginal probabilities. The estimated value of marginal effects on probabilities is also reported in the appendix.

Equation (xii) was initially estimated for general environmental concern, including all explanatory variables. Table III supports the positive impact of higher education and higher income on general environmental concerns. However, we find gender, age, mid-level income, and full employment level are insignificant in the choice of various degrees of environmental concern. Estimated $R^2$ tells that model is correctly predicts 55 % of the choice even though only 4 of 8 variables have significant coefficients. Model, excluding co-linear variables, also validate the social class hypothesis as we find significant positive impact (1.39) of higher education on individual general environmental concern. The marginal effects of each explanatory variable on each concern level are also reported in Table III. A positive value indicates that an increase in the magnitude of the explanatory variable increases the probability that environmental concerns will be of a specific choice. In this regard, the positive impact of higher education is strong only with the category of y = 4 and not for other options. Regarding general environmental concern, marginal effects on the probability of y = 4 indicate that an incremental change in higher education class or an additional year of schooling increases the probability of environmental concern by 48.8% but decreases the probability of choosing "somewhat concern" by 31.8 %. Household size shows a significant negative impact on general environmental concern for both models.

Regression results for the concern about specific environmental issues are presented in Table IV-XI. The findings show that higher education and higher income appear to be consistent with previous literature and serve as a general and significant indicator of specific environmental concern: such as global warming, city areas, and cross-areas pollution, air/water/soil pollution, energy problem, green land, an ecological problem, substances on health, reduction and recycling of waste and living environmental problem such as noise/odor. Females appear to be relatively more concerned about the city areas, and cross-areas pollution than males and are relatively less concerned about environmental considerations when buying electronics items.

Findings support age hypotheses for eight indices, although several results are found to be inconsistent. There is no effect of gender on general environmental concern, global warming, energy problem, air/water/soil pollution, health hazards, and recycling of waste. Age is found significant and shows the negative impact of choosing "concern" in global warming, city areas, and cross-areas pollution, air/water/soil pollution, harmful substances on health, recycling waste, and living environmental problems. At the same time, it is positively related to city green land and ecological problems. Therefore findings support that the younger citizens in Sylhet city are more concerned about various environmental problems than older. Considering employment status, the model with all variables show a positive impact of part-time employment status on the individual general environmental concern while full-time employment status shows mixed scenario and it is found insignificant for six indices. In contrast to the concern about city areas, pollution and air/water/soil pollution, individual with full-time employment status expresses their positive concern about Sylhet city green land and ecological problem. For all models, the average predictive power measured by $R^2$ is 0.53, which is statistically acceptable. The estimated threshold variables are very significant, indicating that the ordered probit model with five different concerns is highly appropriate.

An examination of Table IV- XII indicates that higher education is the dominant explanatory variable with the z statistics ranges from 10.39 to 3.02 (probability value of 0.00), indicating that higher education leads to a greater probability of more concern in environmental protection, i.e., a greater value in y. Another important explanatory variable is the dummy variable income. When income = 0, the respondent is in the lower or mid-level income group; when it equals 1, the respondent is in the higher-income group. For six indices, given the larger z statistic (ranges from 2.38 to 3.84) and small probability value (0.0), it implies that the probability for an individual with higher income, all other factors held constant, to be "concern" about the specific environmental index is greater. The estimated coefficient on gender (dummy variable with the value of 0 for male students and 1 for female respondents) is insignificant for eight indices except for concern about pro-environmental behavior with a z statistic of 2.24. Furthermore, an individual with part-time employment and mid-level income has no significant impact on specific environmental concerns.

The estimated result of attitude towards pro-environmental behavior is presented in Table XII. Concerning individuals' environmental considerations when buying electronics, we find that women are less concerned than their male counterparts. The marginal effects of higher education and full employment status on choice probability are higher( 0.338 and 0.054 when y = 4 and when y = 3 marginal effects on probabilities are 0.026 and 0.005 consecutively ). Age is found significant factor when buying electronics and estimated marginal effects of age on probabilities of choosing "regard its environmental impact as the first issue to consider" and "regard its environmental impact as the second issue to consider" are 0.004 and 0.0005, respectively.

**Conclusion and Policy Implication**

Environmental problems related to urbanizations are mostly local issues, and solutions must be made locally. However, national and multinational environmental research and policy actions should be conducted to spur individual environmental concerns that may be a core to protecting the urban environment. The survey findings reported in this paper represent an initial attempt to understand determinants as policymaking variables involved in environmental concerns. The study suggests a local understanding of individual's values, motives, and pro-environmental behavior that motivate urban environmental concern is needed before we formulate environmental policies or interventions designed to increase pro-environmental behavior. OPM results about the relationship between environmental concerns and socioeconomic determinants, specifically education, income, and individual age, can influence urban policymakers to formulate feasible actions. Exploring the connectivity between income and environmental concerns has brought us closer to understanding the connectivity of City Corporation's taxation policy and financing plan on environmental protection in general. Of the findings this research generated, the most salient is the finding that young with higher education level has an undeniably strong positive correlation with the environmental concern. This work serves as a starting point, and we suggest further study to explore the topic of behavioral aspect environmental concern and willingness to pay to protect city areas environment.

## Appendix

### a. Correlation matrix

| Variables | Female | Age | High_ed | High_inc | Mid_inc | Hsz | Ful_emp | Prt_emp |
|-----------|--------|-----|---------|----------|---------|-----|---------|---------|
| Female | 1.0000 | | | | | | | |
| Age | 0.1600 | 1.0000 | | | | | | |
| High_ed | -0.0402 | 0.2194 | 1.0000 | | | | | |
| High_inc | 0.0352 | -0.0026 | 0.0504 | 1.0000 | | | | |
| Mid_inc | -0.0755 | 0.0500 | -0.1939 | **-0.6421** | 1.0000 | | | |
| Hsz | 0.0487 | 0.0748 | 0.0539 | 0.1241 | -0.0864 | 1.0000 | | |
| Ful_emp | 0.2429 | 0.3764 | 0.1900 | 0.0926 | -0.1433 | -0.0319 | 1.0000 | |
| Prt_emp | -0.1983 | **-0.6047** | -0.1070 | -0.0554 | 0.0201 | -0.0107 | **-0.7752** | 1.0000 |

### b. Ordered Probit Regression(OPM) results

**Table III: general environmental concern (VA)**

| Variables | Coef(se) | Coef(se)** | y=4 | Marginal effects on probabilities y=3 | y=2 | y=1 | y=0 |
|-----------|----------|-----------|-----|------|-----|-----|-----|
| Female | -.017(.15) | -.01(.15) | -.004 | .003 | .000 | .000 | 0.000 |
| Age | .01(.009) | -001(.007) | -.000 | .000 | .000 | .000 | 0.000 |
| High_ed | 1.36*(.16) | 1.39(.15)* | .488 | -.318 | -.081 | -.082 | -0.005 |
| High_inc | .22*(.19) | .13(.14) | .051 | -.035 | -.008 | -.001 | -0.000 |
| Mid_inc | .17(.21) | | | | | | |
| Hsz | -.09*(-2.4) | -.09(.03)* | -.035 | .024 | .005 | .004 | 0.000 |
| Ful_emp | .28(.22) | -.13(.15) | -.050 | .035 | .008 | .007 | 0.000 |
| Prt_emp | .66*(.26) | | | | | | |

** Regression excluding collinear variables. *Significant at 5% level. Log likelihood = -265.14, $R^2$ = 0.55    ; se: standard error

**Table IV: concern about global warming (VB)**

| Variables | Coef(se) | Coef(se)** | y=4 | Marginal effects on probabilities y=3 | y=2 | y=1 | y=0 |
|-----------|----------|-----------|-----|------|-----|-----|-----|
| Female | -.03(.14) | -.014(.14) | -.005 | .002 | .001 | .0012 | .000 |
| Age | -.018*(.00) | -.03(.007)* | -.010 | .005 | .002 | .0023 | .001 |
| High_ed | 1.55*(.15) | 1.60(.15)* | .576 | -.256 | -.097 | -.133 | -.089 |
| High_inc | .07*(.18) | .16(.13) | .065 | -.032 | -.011 | -.014 | -.007 |
| Mid_inc | -.13(.20) | | | | | | |
| Hsz | -.06(-1.79) | -.06(.03) | -.027 | .013 | .004 | .005 | .003 |
| Ful_emp | .10(.21) | -.11(.14) | -.044 | .021 | .007 | .009 | .004 |
| Prt_emp | .36(.25) | | | | | | |

** Regression excluding collinear variables. *Significant at 5% level, **Log likelihood= -324.44, $R^2$ = 0.36**, se: standard error

**Table V : concern about city areas and cross-areas pollution (VC)**

| Variables | Coef(se) | Coef(se)** | Marginal effects on probabilities | | | | |
|---|---|---|---|---|---|---|---|
| | | | y=4 | y=3 | y=2 | y=1 | y=0 |
| Fe male | .25 (.13) | .28 (.13)* | .107 | -.026 | -.042 | -.036 | -.003 |
| Age | -.018*(.008) | -.02(.00)* | -.010 | .002 | .004 | .003 | .000 |
| High_ed | .86*(.14) | .93(.13)* | .349 | -.095 | -.133 | -.111 | -.010 |
| High_inc | .41 (.16) | .49(.12)* | .188 | -.048 | -.073 | -.062 | -.005 |
| Mid_inc | -.13 (.18) | | | | | | |
| Hsz | -.02(.03) | -.02(.03) | -.009 | .002 | .003 | .003 | .0002 |
| Ful_emp | -.11(.20) | -.38(.13)* | -.147 | .038 | .057 | .048 | .004 |
| Prt_emp | .44(.24) | | | | | | |

** Regression excluding collinear variables. *Significant at 5% level , **Log likelihood = -369.78, R²** = 0.49, se: standard error

**Table VI : concern about air/water/soil pollution problem (VD)**

| Variables | Coef(se) | Coef(se)** | Marginal effects on probabilities | | | | |
|---|---|---|---|---|---|---|---|
| | | | y=4 | y=3 | y=2 | y=1 | y=0 |
| Fe male | -.04 (.14) | -.02(.14) | -.010 | .006 | .002 | .001 | .000 |
| Age | -.01 (.008) | -.02(.00)* | -.008 | .004 | .001 | .001 | .000 |
| High_ed | .79*(.14) | .83(.14)* | .321 | -.081 | -.064 | -.053 | -.014 |
| High_inc | .27(.17) | .35(.13)* | .138 | -.081 | -.028 | -.023 | -.005 |
| Mid_inc | -.11(.19) | | | | | | |
| Hsz | -.06 (.03) | -.06(.03) | -.027 | .016 | .005 | .004 | .001 |
| Ful_emp | -.37 (.22) | -.46(.14)* | -.182 | .107 | .037 | .030 | .007 |
| Prt_emp | .15(.26) | | | | | | |

** Regression excluding collinear variables. *Significant at 5% level, **Log likelihood = -307.05, R² = 0.57**, se: standard error

**Table VII: concern about city areas energy problem (VE)**

| Variables | Coef(se) | Coef(se) | Marginal effects on probabilities | | | | |
|---|---|---|---|---|---|---|---|
| | | | y=4 | y=3 | y=2 | y=1 | y=0 |
| Female | -.13(.13) | -.12(.13) | -.049 | .010 | .014 | .016 | .007 |
| Age | -.02*(.008) | -.02(.00)* | -.009 | .001 | .002 | .003 | .001 |
| High_ed | .74*(.13) | .76(.13)* | -.285 | -.061 | .080 | -.098 | -.045 |
| High_inc | .17(.16) | .24(.12) | -.091 | -.018 | -.026 | -.032 | -.014 |
| Mid_inc | -.13(.18) | | | | | | |
| Hsz | -.003(.035) | -.002(.03) | -.001 | .000 | .000 | .000 | .000 |
| Ful_emp | .06(.20) | .07(.13) | .028 | -.005 | -.008 | -.009 | -.004 |
| Prt_emp | -.001(.24) | | | | | | |

** Regression excluding collinear variables. *Significant at 5% level, **Log likelihood = -398.73, R² = 0.58**, se: standard error

**Table VIII : concern about Sylhet city green land and ecological problem (VF),**

| Variables | Coef(se) | Coef(se) | Marginal effects on probabilities | | | | |
|---|---|---|---|---|---|---|---|
| | | | y=4 | y=3 | y=2 | y=1 | y=0 |
| Female | .14(.13) | .18(.13) | .067 | -.007 | -.015 | -.037 | -.007 |
| Age | -.006(.008) | .01(.00)* | -.006 | .000 | .001 | .003 | .000 |
| High_ed | .89*(.13) | .98(.13)* | .359 | -.050 | -.079 | -.192 | -.043 |
| High_inc | -.18(.16) | .05(.12) | .021 | -.002 | -.004 | -.011 | -.002 |
| Mid_inc | -.39*(.18) | | | | | | |
| Hsz | -.008(.03) | -.01(.03) | -.003 | .000 | .000 | .002 | .000 |
| Ful_emp | .50*(.20) | .28(.13)* | .106 | -.013 | -.024 | -.057 | -.011 |
| Prt_emp | .38(.23) | | | | | | |

** Regression excluding collinear variables. *Significant at 5% level, **Log likelihood = -405.78, R² = 0.68**, se: standard error

**Table IX: concern about the effect of harmful substances on health (VG)**

------------------------------------------------------------------------------------------------------

| Variables | Coef(se) | Coef(se) | | Marginal effects on probabilities | | | |
| | | | y=4 | y=3 | y=2 | y=1 | y=0 |
|-----------|----------|----------|-----|-----|-----|-----|-----|
| Female | .15(.13) | .16(.12)* | .054 | .006 | -.026 | -.020 | -.013 |
| Age | -.02*(.008) | -.02(.006)* | -.008 | -.000 | .004 | .003 | .002 |
| High_ed | .53*(.13) | .58(.12)* | .191 | .012 | -.089 | -.068 | -.046 |
| High_inc | .15(.16) | .29(.12)* | .095 | .009 | -.045 | -.035 | -.023 |
| Mid_inc | -.26(.18) | | | | | | |
| Hsz | -.04(.03) | -.04(.03) | -.014 | -.001 | .006 | .005 | .003 |
| Ful_emp | -.01(.20) | -.10(.13) | -.034 | -.003 | .016 | .012 | .008 |
| Prt_emp | .16(.23) | | | | | | |

** Regression excluding collinear variables.  *Significant at 5% level,  **Log likelihood = -424.78,  R$^2$ = 0.46**, se: standard error

**Table X: concern about disposal, reduction, and  recycling of waste (VH)**

------------------------------------------------------------------------------------------------------

| Variables | Coef(se) | Coef(se) | | Marginal effects on probabilities | | | |
| | | | y=4 | y=3 | y=2 | y=1 | y=0 |
|-----------|----------|----------|-----|-----|-----|-----|-----|
| Female | -.015(.12) | -.007(.12) | -.002 | -.000 | .000 | .001 | .000 |
| Age | -.012(.008) | -.016(.006)* | -.004 | -.002 | .001 | .002 | .001 |
| High_ed | .69*(.13) | .72(.12)* | .194 | .079 | -.070 | -.121 | -.082 |
| High_inc | .04(.15) | .13(.12) | .034 | .016 | -.013 | -.022 | -.015 |
| Mid_inc | -.15(.18) | | | | | | |
| Hsz | -.02(.03) | -.02(.03) | -.006 | -.003 | .002 | .004 | .002 |
| Ful_emp | .10(.20) | .03(.04) | .009 | .004 | -.003 | -.006 | -.004 |
| Prt_emp | .11(.23) | | | | | | |

** Regression excluding collinear variables.  *Significant at 5% level,  **Log likelihood = -453.26,  R$^2$ = 0.64**, se: standard error

**Table XI: concern about living environmental problems such as noise/odor (VI)**

------------------------------------------------------------------------------------------------------

| Variables | Coef(se) | Coef(se) | | Marginal effects on probabilities | | | |
| | | | y=4 | y=3 | y=2 | y=1 | y=0 |
|-----------|----------|----------|-----|-----|-----|-----|-----|
| Female | .15(.13) | .13(.13) | .051 | -.021 | -.017 | -.009 | -.002 |
| Age | -.014(.008) | -.01(.007)* | -.006 | .002 | .002 | .001 | .000 |
| High_ed | .57*(.14) | .54(.13)* | .213 | -.093 | -.072 | -.038 | -.008 |
| High_inc | .47*(.16) | .35(.12)* | .137 | -.058 | -.047 | -.025 | -.005 |
| Mid_inc | .18(.18) | | | | | | |
| Hsz | -.08*(.036) | -.08(.03)* | -.034 | .015 | .011 | .006 | .001 |
| Ful_emp | -.16(.21) | -.25(.14) | -.100 | .043 | .034 | .018 | .004 |
| Prt_emp | .12(.25) | | | | | | |

------------------------------------------------------------------------------------------------------

** Regression excluding collinear variables.  *Significant at 5% level,  **Log likelihood = -343.09,  R$^2$ = 0.49**, se: standard error

**Table XII: environmental Considerations when buying electronics (VJ).**

------------------------------------------------------------------------------------------------------

| Variables | Coef(se) | Coef(se) | | Marginal effects on probabilities | | | |
| | | | y=4 | y=3 | y=2 | y=1 | y=0 |
|-----------|----------|----------|-----|-----|-----|-----|-----|
| Female | -.33*(.13) | -.30(.13)* | -.110 | -.008 | .014 | .014 | .090 |
| Age | .02*(.008) | .01(.006) | .004 | .000 | -.000 | -.000 | -.003 |
| High_ed | .87*(.13) | .95(.13)* | .338 | .026 | -.039 | -.041 | -.283 |
| High_inc | -.16(.17) | -.03(.12) | -.014 | -.001 | .001 | .001 | .012 |
| Mid_inc | -.18(.19) | | | | | | |
| Hsz | -.003(.036) | -.005(.03) | -.002 | -.000 | .000 | .000 | .001 |
| Ful_emp | .51*(.20) | .14(.14) | .054 | .005 | -.006 | -.006 | -.046 |
| Prt_emp | .61*(.24) | | | | | | |

------------------------------------------------------------------------------------------------------

** Regression excluding collinear variables.  *Significant at 5% level,  **Log likelihood = -343.09,  R$^2$ = 0.74**, se: standard error